\documentclass[seceq,preprint]{ptptex}
\usepackage{wrapft}
\usepackage{graphicx}

\newcommand{\bbra}[1]{\langle\!\langle {#1} |}     
\newcommand{\kket}[1]{| {#1} \rangle\!\rangle}     
\newcommand{\wtilde}[1]{\widetilde{#1}} 

\newcommand{\kbar}{k \kern -0.5em\raise 0.6ex \hbox{--}}

\def\<{\langle}
\def\>{\rangle}
\def\bsub{\begin{subequations}}
\def\esub{\end{subequations}}
\def\beqn{\begin{eqnarray}}
\def\eeqn{\end{eqnarray}}
\def\beq{\begin{equation}}
\def\eeq{\end{equation}}
\def\b{\begin{equation}}

\title{
Classical and Quantal Descriptions of Small Amplitude Fluctuations Around 
Equilibriums in the Two-Level Pairing Model
}

\author{
Yasuhiko {\sc Tsue},$^{1}$ 
Constan\c{c}a {\sc Provid\^encia},$^{2}$\\
Jo\~ao da {\sc Provid\^encia}$^{2}$ 
and Masatoshi {\sc Yamamura}$^{3}$
}

\inst{
$^1$Physics Division, Faculty of Science, Kochi University, Kochi 780-8520, 
Japan\\
$^{2}$Departamento de F\'\i sica, Universidade de Coimbra, P-3004-516 Coimbra, 
Portugal\\
$^{3}$Faculty of Engineering, Kansai University, Suita 564-8680, Japan
}

\recdate{
\today
}

\abst{
Various classical counterparts for the two-level pairing model in a 
many-fermion system are presented in the Schwinger boson representation. 
It is shown that 
one of the key ingredients giving the classical descriptions for 
quantal system is the use of the 
various trial states besides the $su(2)\otimes su(2)$-coherent state, 
which may be natural selection for the two-level pairing model governed by 
the $su(2)\otimes su(2)$-algebra. 
It is pointed out that 
the fictitious behavior like the sharp phase transition can be avoided by 
using the other states such as the $su(2)\otimes su(1,1)$- and 
the $su(1,1)\otimes su(1,1)$-coherent states, while the sharp 
phase transition appears in the usual Hartree-Fock-Bogoliubov and 
the quasi-particle random phase approximations in the original 
fermion system. 
}
\begin{document}
\maketitle

\section{Introduction}

The use of boson representation for many-fermion systems gives a powerful 
method to describe the dynamics of the original fermion systems. 
On the basis of the Lie algebraic structure of the model Hamiltonian, 
the Schwinger \cite{S} and the Holstein-Primakoff \cite{HP} type 
boson realizations give 
helpful tools to investigate the dynamics of the original fermion 
systems. 
The boson mapping method, which was proposed in Refs.\citen{BZ}, \citen{MYT} 
and \citen{P}, is also useful method \cite{KM} 
to describe the many-body systems 
beyond the usual mean field, 
the Hartree-Fock(-Bogoliubov) and the random phase approximation. 
The Marumori-Yamamura-Tokunaga (MYT) boson mapping method \cite{MYT} 
is characterized by the mapping of operators acting on the Fock space 
in the original systems into operators acting on the boson Fock space 
in the corresponding boson systems. 
This idea was strongly stressed by the present authors and widely applied 
to the many-fermion and many-boson systems 
to study the dynamics under consideration.\cite{suppl}
The oscillation around the mean field configuration is, of course, 
described in terms of the bosonic degrees of freedom. 

As another advantage of the use of the 
boson representation of the original fermion 
system, we can say that 
it is possible to introduce various trial states in the framework 
of the time-dependent variational method. 
In the fermion system, the trial state may be restricted to 
a simple class in the mean field approximation. 
For example, if the fermion 
system is governed by the $su(2)\otimes su(2)$-algebra constructed 
from the bilinear forms of the fermion creation and annihilation 
operators, the trial state in the variational method may be restricted to the 
Slater determinantal or BCS state, namely, the $su(2)\otimes su(2)$-coherent 
state in the framework of the mean field 
approximation. 
This is just the case in which the two-level pairing model in a 
many-fermion system is considered. 
On the other hand, in the boson representation of the original fermion system, 
various trial states are possible, even if the system is governed by, 
for example, the $su(2)\otimes su(2)$-algebra. 
The $su(2)\otimes su(2)$-coherent state corresponds to the Glauber coherent 
state in the boson representation. In addition to this state, the 
$su(2)\otimes su(1,1)$- and the $su(1,1)\otimes su(1,1)$-coherent states 
are possible to describe the dynamics of the original two-level pairing model 
governed by the $su(2)\otimes su(2)$-algebra. 
Thus, it is possible to give the various classical descriptions for 
the original unique many-fermion system.

In this paper, we deal with the two-level pairing model governed by the 
$su(2)\otimes su(2)$-algebra. This model is treated in the Schwinger 
boson representation for the $su(2)\otimes su(2)$-algebra in which 
four kinds of boson operators are introduced. 
In order to give a classical description of the pairing model, 
three types of trial states are introduced, namely, the 
$su(2)\otimes su(2)$-, $su(2)\otimes su(1,1)$- and 
$su(1,1)\otimes su(1,1)$-coherent states. 
The $su(2)\otimes su(1,1)$-coherent state for the two-level pairing model 
has been defined in Ref. \citen{1}, which is referred to as (I). 
Also, the 
$su(1,1)\otimes su(1,1)$-coherent state for the two-level pairing model 
has been defined in Ref.\citen{1-1}. 
In this paper, we numerically 
calculate the ground state energy and the frequency 
of the small amplitude oscillation around the energy minimum state, and 
compare them with the exact results. 
Then, it is pointed out that 
the fictitious behavior like the sharp phase transition can be avoided by 
using the $su(2)\otimes su(1,1)$- and 
the $su(1,1)\otimes su(1,1)$-coherent states, while the sharp 
phase transition appears by using of the $su(2)\otimes su(2)$-coherent 
state which corresponds to the usual Hartree-Fock-Bogoliubov and 
the quasi-particle random phase approximations in the original 
fermion system. 
As an additional remark, it should be mentioned that, from a viewpoint 
different from the present one, the authors have already reported numerical 
results for the two-level pairing model, which supplement the 
present results.\cite{JPhys}

This paper is organized as follows. 
The outline of the two-level pairing model and its Schwinger boson 
representation with the four kinds of boson operators 
is presented in the next section. 
In \S 3, re-formation of this model 
in terms of one kind of boson operator and its 
classical counterpart is given and three types of boson coherent states 
are introduced. 
A classical description based on each one of the three types of 
coherent state is 
given in \S 4. 
In order to investigate the ground state energy with quantum correction 
and 
the frequency of the small amplitude oscillation around the static 
configuration, the quantal treatment is given in \S 5. 
The numerical results are also shown in this section. 
The last section is devoted to concluding remarks. 
In appendix A, some results in the case of the $su(2)\otimes su(2)$-coherent 
state are supplemented. 
In appendix B, the derivation of quantal Hamiltonian is given 
in order to introduce the ground state energy with quantum correction and 
the frequency of small amplitude oscillation around the ground state.

\section{Outline of the model}

In this section and partly in the next one, we list some of the 
relations appearing in (I). 
They help us to understand various results shown in this paper. 
The Hamiltonian of the system which we intend to describe is 
given in the relation (I$\cdot$3$\cdot$3a): 
\begin{equation}\label{2-1}
{\tilde H}=\epsilon \left({\wtilde S}_0(+)-{\wtilde S}_0(-)\right)
-G\left({\wtilde S}_+(+)+{\wtilde S}_+(-)\right)
\left({\wtilde S}_-(+)+{\wtilde S}_-(-)\right) \ .
\end{equation}
Here, $\epsilon$ and $G$ denote the energy difference between the upper 
(specified by $\sigma=+$) and the lower (specified by $\sigma=-$) level 
and the strength of the interaction, respectively. 
The set $({\wtilde S}_{\pm,0}(\sigma);\ \sigma=\pm)$ obeys the 
$su(2)$-algebra and it can be expressed in the form shown in the relation 
(I$\cdot$3$\cdot$1): 
\begin{equation}\label{2-2}
{\wtilde S}_+(\sigma)=\hbar{\hat a}_\sigma^*{\hat b}_\sigma\ , \qquad
{\wtilde S}_-(\sigma)=\hbar{\hat b}_\sigma^*{\hat a}_\sigma\ , \qquad
{\wtilde S}_0(\sigma)=(\hbar/2)({\hat a}_\sigma^*{\hat a}_\sigma 
-{\hat b}_\sigma^*{\hat b}_\sigma) \ .
\end{equation}
The operators $({\hat a}_{\sigma},{\hat a}_{\sigma}^*)$ and 
$({\hat b}_{\sigma},{\hat b}_{\sigma}^*)$ denote boson annihilation and 
creation 
operators. 
The form (\ref{2-2}) is identical to the Schwinger boson representation of the 
$su(2)$-algebra. 
From the composition of the Hamiltonian (\ref{2-1}), we can understand that 
the present system obeys the $su(2)\otimes su(2)$-algebra. 
The original form of the Hamiltonian (\ref{2-1}) can be expressed in terms of 
fermion operators shown in the form (I$\cdot$2$\cdot$1). 
The explicit expression of the Hamiltonian (\ref{2-1}) in boson operators 
is given in the relation (I$\cdot$3$\cdot$3b). 

For the present boson system, the following four hermitian operators are 
mutually commutable: 
\bsub\label{2-3}
\beqn
& &{\wtilde L}=(\hbar/2)({\hat a}_+^*{\hat a}_+ + {\hat b}_+^*{\hat b}_+) \ , 
\label{2-3a}\\
& &{\wtilde M}=(\hbar/2)({\hat a}_+^*{\hat a}_+ + {\hat a}_-^*{\hat a}_-) \ , 
\label{2-3b}\\
& &{\wtilde T}=(\hbar/2)(-{\hat a}_+^*{\hat a}_+ + {\hat b}_-^*{\hat b}_-
+1) \ , 
\label{2-3c}\\
& &{\wtilde K}=\hbar{\hat a}_+^*{\hat a}_+  \ . 
\label{2-3d}
\eeqn
\esub
The form (\ref{2-3}) is shown in the relation (I$\cdot$3$\cdot$4) and 
it should be noted that ${\wtilde L}$, ${\wtilde M}$ and ${\wtilde T}$ 
commute with the Hamiltonian (\ref{2-1}). 
With the use of the operators (\ref{2-3}), together with the expression 
(\ref{2-2}), the Hamiltonian (\ref{2-1}) can be rewritten as 
\begin{equation}\label{2-4}
{\wtilde H}={\wtilde H}_0+{\wtilde H}_1 \ , 
\qquad\qquad\qquad\qquad\qquad\qquad\qquad\qquad\qquad
\end{equation}
\vspace{-0.5cm}
\bsub\label{2-5}
\beqn
& &{\wtilde H}_0=-\left[\epsilon\left({\wtilde L}+{\wtilde M}-
\left(\wtilde T-\hbar/2\right)\right)+4G{\wtilde T}{\wtilde M}\right] 
\nonumber\\
& &\qquad\quad
+2\left[\epsilon-G\left({\wtilde L}+{\wtilde M}-
\left(\wtilde T-\hbar/2\right)\right)\right]{\wtilde K} +2G{\wtilde K}^2 \ , 
\label{2-5a}\\
& &{\wtilde H}_1=-G\cdot \hbar^2\left[
{\hat a}_+^*{\hat b}_-^*{\hat b}_+{\hat a}_-
+{\hat a}_-^*{\hat b}_+^*{\hat b}_-{\hat a}_+ \right] \ . 
\label{2-5b}
\eeqn
\esub
In (I), noting that ${\wtilde L}$, ${\wtilde M}$ and ${\wtilde T}$ are 
constants of motion, we treated the Hamiltonian (\ref{2-4}) 
quantum- and classical-mechanically. 
In the case of the classical treatment in (I), 
the $su(2)\otimes su(1,1)$-coherent state played a central role. 
In this paper, we treat the cases of the $su(1,1)\otimes su(1,1)$- 
and the $su(2)\otimes su(2)$-coherent states on an equal footing with 
the $su(2)\otimes su(1,1)$-coherent state.

As was mentioned in (I), the two-level pairing model as a many-fermion system 
is characterized by three quantities. 
They are the numbers of the single-particle states in the levels 
$\sigma=+$ and $\sigma=-$, namely, $2\Omega_+$ and $2\Omega_-$, respectively, 
and the total fermion number $N$. 
The original fermion system is related to the present boson system through 
\begin{equation}\label{2-6}
(\hbar/2)\Omega_+=L \ , \qquad
(\hbar/2)N=M \ , \qquad
(\hbar/2)(\Omega_- - N/2 +1)=T \ . 
\end{equation}
Here, the $q$-numbers ${\wtilde L}$, ${\wtilde M}$ and ${\wtilde T}$ are 
replaced with the $c$-numbers $L$, $M$ and $T$, respectively. 
Through the relation (\ref{2-6}), the subspace, in which the eigenvalues of 
${\wtilde L}$, ${\wtilde M}$ and ${\wtilde T}$ are $L$, $M$ and $T$, 
respectively, corresponds to the original fermion space specified 
by $\Omega_+$, $\Omega_-$ and $N$. 
The above can be found in the relation (I$\cdot$6$\cdot$3). 
In this paper, we treat the case $T\geq \hbar/2$. 
For the numerical results, we discuss the case 
\begin{equation}\label{2-7}
\Omega_+=\Omega_- \ (=\Omega)\ , \qquad N=2\Omega\ .
\end{equation}
The case (\ref{2-7}) corresponds to 
\begin{equation}\label{2-8}
L=M=(\hbar/2)\Omega \ , \qquad T=\hbar/2 \ .
\end{equation}
The case (\ref{2-7}) tells that the degeneracies of the two levels 
are the same and the total fermions occupy the lower level completely 
(a closed shell), if the interaction is switched off.

\section{Re-formation in terms of one kind of boson operator and 
its classical counterpart}

The system presented in \S 2 can be expressed in terms of four kinds of 
bosons and it contains three constants of motion. 
Therefore, the present system can be described in terms of 
one kind of degree of freedom. 
With the aid of the MYT mapping method, in (I), we transcribed the system 
in the space spanned by one kind of boson $({\hat c}, {\hat c}^*)$. 
The Hamiltonian (\ref{2-4}) is transcribed in the form 
\begin{equation}\label{3-1}
{\hat H}={\hat H}_0+{\hat H}_1 \ , 
\qquad\qquad\qquad\qquad\qquad\qquad\qquad\qquad\qquad\qquad\ \ 
\end{equation}
\vspace{-0.5cm}
\bsub\label{3-2}
\beqn
& &{\hat H}_0=-\left[\epsilon\left({L}+{M}-\left(T-\hbar/2\right)\right)
+4G{T}{M}\right] 
\nonumber\\
& &\qquad\quad
+2\left[\epsilon-G\left({L}+{M}-\left(T-\hbar/2\right)\right)\right]
{\hat K} +2G{\hat K}^2 \ , 
\label{3-2a}\\
& &{\hat H}_1={\hat H}_1(c_-)={\hat H}_1(c^0)={\hat H}_1(c) \ , \nonumber\\
& &\ {\hat H}_1(c_-)=
-G\biggl[\sqrt{\hbar}{\hat c}^*\cdot\sqrt{2T+{\hat K}}\sqrt{2M-{\hat K}}
\sqrt{2L-{\hat K}} \nonumber\\
& &\qquad\qquad\qquad\ \ 
+\sqrt{2L-{\hat K}}\sqrt{2M-{\hat K}}\sqrt{2T+{\hat K}}\cdot \sqrt{\hbar}\ 
{\hat c}\biggl] \ , \nonumber\\
& &\ {\hat H}_1(c^0)=
-G\biggl[\sqrt{2L+\hbar-{\hat K}}\cdot\sqrt{\hbar}{\hat c}^*
\cdot\sqrt{2T+{\hat K}}\sqrt{2M-{\hat K}}
\nonumber\\
& &\qquad\qquad\qquad\ \ 
+\sqrt{2M-{\hat K}}\sqrt{2T+{\hat K}}\cdot \sqrt{\hbar}\ {\hat c}\cdot
\sqrt{2L+\hbar-{\hat K}}\biggl] \ , \nonumber\\
& &\ {\hat H}_1(c)=
-G\biggl[\sqrt{2T-\hbar+{\hat K}}\cdot\sqrt{\hbar}{\hat c}^*
\cdot\sqrt{2M-{\hat K}}\sqrt{2L-{\hat K}}
\nonumber\\
& &\qquad\qquad\qquad\ \ 
+\sqrt{2L-{\hat K}}\sqrt{2M-{\hat K}}\cdot \sqrt{\hbar}\ {\hat c}\cdot
\sqrt{2T-\hbar+{\hat K}}\biggl] \ ,  
\label{3-2b}
\eeqn
\esub
\vspace{-0.5cm}
\begin{equation}\label{3-3}
{\hat K}=\hbar{\hat c}^*{\hat c} \ . 
\qquad\qquad\qquad\qquad\qquad\qquad\qquad\qquad\qquad
\end{equation}
The notations $c_-$, $c^0$ and $c$ will be interpreted later. 
The relation 
$[\sqrt{\hbar}{\hat c}^*\ , \ {\hat K}]=-\hbar\cdot
\sqrt{\hbar}{\hat c}^*$ and 
$[\sqrt{\hbar}{\hat c}\ , \ {\hat K}]=\hbar\cdot
\sqrt{\hbar}{\hat c}$ shows that ${\hat H}_1(c_-)
={\hat H}_1(c^0)={\hat H}(c)$. 
The above can be seen in the relations (I$\cdot$4$\cdot$11) and 
(I$\cdot$4$\cdot$12) with the derivation.

We know that, with the aid of the appropriately chosen boson coherent state, 
we can derive the classical counterpart of the original quantal system. 
For the present system, we used in (I), the coherent state 
$\kket{c_-}$ shown in the form (I$\cdot$5$\cdot$3) and its rewritten form 
(I$\cdot$5$\cdot$6). 
The state $\kket{c_-}$ can be, further, rewritten as 
\beqn\label{3-4}
\kket{c_-}&=&
N_c\exp\left(\frac{V_+}{U_+}{\hat a}_+^*{\hat b}_-^*\right)
\exp\left(\frac{1}{\sqrt{\hbar}}\frac{W_-}{U_+}{\hat b}_-^*\right)
\nonumber\\
& &\times \exp\left(\frac{A_-}{B_+}{\hat a}_-^*{\hat b}_+\right)
\exp\left(\frac{1}{\sqrt{\hbar}}B_+{\hat b}_+^*\right)\kket{0} \ , 
\nonumber\\
U_+&=&\sqrt{1+|V_+|^2} \ . 
\eeqn
Here, $N_c$ denotes the normalization constant and $V_+$, $W_-$, 
$A_-$ and $B_+$ are complex parameters. 
We can see that the state $\kket{c_-}$ is generated by successive operation 
of ${\hat a}_-^*{\hat b}_+$ and ${\hat a}_+^*{\hat b}_-^*$. 
The operators ${\hat a}_-^*{\hat b}_+$ and ${\hat a}_+^*{\hat b}_-^*$ are 
the raising operators of the $su(2)$- and the $su(1,1)$-algebra, 
respectively. 
From the above reason, in (I), we called the state $\kket{c_-}$ the 
$su(2)\otimes su(1,1)$-coherent state. 
Through the process presented in (I), the expectation value of the 
Hamiltonian ${\wtilde H}$, which is given in the relation 
(I$\cdot$3$\cdot$3), was calculated. 
The result is shown in the relation (I$\cdot$4$\cdot$15). 
In notations slightly different from those in (I), we have the 
following form: 
\begin{equation}\label{3-5}
H(c_-)=\bbra{c_-}{\wtilde H} \kket{c_-}=H_0(c_-)+H_1(c_-) \ , 
\qquad\qquad\qquad\ \ 
\end{equation}
\vspace{-0.5cm}
\bsub\label{3-6}
\beqn
& &H_0(c_-)=-\left[\epsilon\left({L}+{M}-\left(T-\hbar/2\right)\right)
+4G{T}{M}\right] 
\nonumber\\
& &\qquad\quad\qquad
+2\left[\epsilon-G\left({L}+{M}-\left(T-\hbar/2\right)\right)\right]
{K} +2G{K}^2 \ , 
\label{3-6a}\\
& &H_1(c_-)=-2G\sqrt{K(2T+K)(2M-K)(2L-K)}\cos \psi \ . 
\label{3-6b}
\eeqn
\esub
Here, $(\psi, K)$ denotes the angle-action variable.

In (I), we showed various forms of coherent states. 
One of them was presented in the form (I$\cdot$7$\cdot$20), which is 
denoted as $\kket{c^0}$. 
The state $\kket{c^0}$ can be rewritten as 
\beqn\label{3-7}
\kket{c^0}&=&
N_c\exp\left(\frac{V_+}{U_+}{\hat a}_+^*{\hat b}_-^*\right)
\exp\left(\frac{1}{\sqrt{\hbar}}\frac{W_-}{U_+}{\hat b}_-^*\right)
\nonumber\\
& &\times \exp\left(\frac{V_-}{U_-}{\hat a}_-^*{\hat b}_+^*\right)
\exp\left(\frac{1}{\sqrt{\hbar}}\frac{W_+}{U_-}{\hat b}_+^*\right)\kket{0} \ , 
\nonumber\\
U_\sigma&=&\sqrt{1+|V_\sigma|^2} \ . 
\eeqn
Here, of course, $N_c$ denotes a normalization constant and 
$V_{\sigma}$ and $W_{\sigma}$ ($\sigma=\pm)$ are complex parameters. 
We can see that $\kket{c^0}$ is generated by the operators 
${\hat a}_+^*{\hat b}_-^*$ and ${\hat a}_-^*{\hat b}_+^*$, which are 
the raising operators of two independent $su(1,1)$-algebras. 
Then, we call the state (\ref{3-7}) the $su(1,1)\otimes su(1,1)$-coherent 
state. 
Under the same idea as that in the case of $\kket{c_-}$, the expectation value 
of ${\wtilde H}$ for $\kket{c^0}$ is given in the form 
\begin{equation}\label{3-8}
H(c^0)=\bbra{c^0}{\wtilde H} \kket{c^0}=H_0(c^0)+H_1(c^0) \ , 
\qquad\qquad\qquad\qquad\ \ 
\end{equation}
\vspace{-0.5cm}
\bsub\label{3-9}
\beqn
& &H_0(c^0)=H_0(c_-) \ , 
\label{3-9a}\\
& &H_1(c^0)=-2G\sqrt{K(2T+K)(2M-K)(2L+\hbar-K)}\cos \psi \ . 
\label{3-9b}
\eeqn
\esub
The third is the following one: 
\beqn\label{3-10}
\kket{c}&=&
N_c\exp\left(\frac{A_+}{B_-}{\hat a}_+^*{\hat b}_-\right)
\exp\left(\frac{1}{\sqrt{\hbar}}B_-{\hat b}_-^*\right)
\nonumber\\
& &\times \exp\left(\frac{A_-}{B_+}{\hat a}_-^*{\hat b}_+\right)
\exp\left(\frac{1}{\sqrt{\hbar}}B_+{\hat b}_+^*\right)\kket{0} \ . 
\eeqn
Here, $N_c$ denotes the normalization constant and $A_\sigma$ and 
$B_\sigma$ $(\sigma=\pm)$ are complex parameters. 
Clearly, $\kket{c}$ is generated by ${\hat a}_+^*{\hat b}_-$ and 
${\hat a}_-^*{\hat b}_+$, which are the raising operators 
of two independent $su(2)$-algebras. 
Therefore, we call it the $su(2)\otimes su(2)$-coherent state. 
We did not contact with $\kket{c}$ explicitly in (I). 
The state $\kket{c}$ can be rewritten as 
\beqn\label{3-11}
\kket{c}&=&
N_c\exp\left(\frac{A_+}{B_+}{\hat a}_+^*{\hat b}_+\right)
\exp\left(\frac{B_+}{\sqrt{\hbar}}{\hat b}_+^*\right)
\nonumber\\
& &\times \exp\left(\frac{A_-}{B_-}{\hat a}_-^*{\hat b}_-\right)
\exp\left(\frac{B_-}{\sqrt{\hbar}}{\hat b}_-^*\right)\kket{0} \nonumber\\
&=&N_c\exp\left(\frac{1}{\sqrt{\hbar}}\left(
A_+{\hat a}_+^* + B_+{\hat b}_+^* + A_-{\hat a}_-^* + B_-{\hat b}_-^*\right)
\right)\kket{0}\ . 
\eeqn
The form (\ref{3-11}) tells that $\kket{c}$ corresponds to the BCS state 
and it is identical to a kind of the Glauber coherent state. 
Then, the expectation value of ${\wtilde H}$ for $\kket{c}$ is calculated 
in the form 
\begin{equation}\label{3-12}
H(c)=\bbra{c}{\wtilde H} \kket{c}=H_0(c)+H_1(c) \ , 
\qquad\qquad\qquad\qquad\qquad\ \ 
\end{equation}
\vspace{-0.5cm}
\bsub\label{3-13}
\beqn
& &H_0(c)=H_0(c_-) \ , 
\label{3-13a}\\
& &H_1(c)=-2G\sqrt{K(2T-\hbar+K)(2M-K)(2L-K)}\cos \psi \ . 
\label{3-13b}
\eeqn
\esub

In the above, we showed that, for one quantal system, three classical 
counterparts were derived. 
It may be natural, because we used three different coherent states. 
Our final problem of this section is related to the re-quantization. 
For this task, it may be convenient to introduce new canonical 
variable in boson-type, $(c,c^*)$, through 
\begin{equation}\label{3-14}
\sqrt{\hbar}c=\sqrt{K}e^{-i\psi} \ , \qquad 
\sqrt{\hbar}c^*=\sqrt{K}e^{i\psi}\ . \qquad
(K=\hbar c^*c)
\end{equation}
The re-quantization may be performed by the replacement 
\begin{equation}\label{3-15}
\sqrt{\hbar}c \longrightarrow \sqrt{\hbar}{\hat c} \ , \qquad
\sqrt{\hbar}c^* \longrightarrow \sqrt{\hbar}{\hat c}^* \ , \qquad
K \longrightarrow {\hat K} \ .
\end{equation}
Under the replacement (\ref{3-15}), $H_0(c_-)=H_0(c^0)=H_0(c)$ is 
requantized to ${\hat H}_0$ shown in the relation (\ref{3-2a}). 
Concerning $H_1(c_-)$, $H_1(c^0)$ and $H_1(c)$, the following 
relation is interesting: 
\bsub\label{3-16}
\beqn
& &\sqrt{\hbar}c^*\sqrt{2T+K}\sqrt{2M-K}\sqrt{2L-K}
\longrightarrow 
\sqrt{\hbar}{\hat c}^*\sqrt{2T+{\hat K}}\sqrt{2M-{\hat K}}\sqrt{2L-{\hat K}}
\nonumber\\
& &\qquad\qquad\qquad\qquad\qquad\qquad\qquad\qquad\qquad\qquad\qquad\qquad
{\rm for}\quad\ \kket{c_-}
\label{3-16a}\\
& &\sqrt{\hbar}c^*\sqrt{2T+K}\sqrt{2M-K}\sqrt{2L+\hbar-K}
=\sqrt{2L+\hbar-K}\sqrt{\hbar}c^*\sqrt{2T+K}\sqrt{2M-K}\nonumber\\
& &\longrightarrow 
\sqrt{2L+\hbar-{\hat K}}\sqrt{\hbar}{\hat c}^*\sqrt{2T+{\hat K}}
\sqrt{2M-{\hat K}}=
\sqrt{\hbar}{\hat c}^*\sqrt{2T+{\hat K}}\sqrt{2M-{\hat K}}\sqrt{2L-{\hat K}}
\nonumber\\
& &\qquad\qquad\qquad\qquad\qquad\qquad\qquad\qquad\qquad\qquad\qquad\qquad
{\rm for}\quad\ \kket{c^0}
\label{3-16b}\\
& &\sqrt{\hbar}c^*\sqrt{2T-\hbar+K}\sqrt{2M-K}\sqrt{2L-K}
=\sqrt{2T-\hbar+K}\sqrt{\hbar}c^*\sqrt{2M-K}\sqrt{2L-K}\nonumber\\
& &\longrightarrow 
\sqrt{2T-\hbar+{\hat K}}\sqrt{\hbar}{\hat c}^*\sqrt{2M-{\hat K}}
\sqrt{2L-{\hat K}}=
\sqrt{\hbar}{\hat c}^*\sqrt{2T+{\hat K}}\sqrt{2M-{\hat K}}\sqrt{2L-{\hat K}}
\nonumber\\
& &\qquad\qquad\qquad\qquad\qquad\qquad\qquad\qquad\qquad\qquad\qquad\qquad
{\rm for}\quad\ \kket{c}
\label{3-16c}
\eeqn
\esub
If we note the relation $2K\cos \psi=\sqrt{K}(\sqrt{\hbar}c^*+\sqrt{\hbar}c)$, 
the relation (\ref{3-16}) leads to 
\begin{equation}\label{3-17}
H_1(c_-)\longrightarrow {\hat H}_1(c_-) \ , \qquad
H_1(c^0)\longrightarrow {\hat H}_1(c^0) \ , \qquad
H_1(c)\longrightarrow {\hat H}_1(c) \ . 
\end{equation}
As was shown in the relation (\ref{3-2b}), we have ${\hat H}_1(c_-)=
{\hat H}_1(c^0)={\hat H}_1(c)={\hat H}_1$. 
In the above, the meanings of the symbols $c_-$, $c^0$ and $c$ may be clear. 
From the above argument, we can understand that three different 
classical systems become to single quantal system, 
depending on the ordering of the variables. 
Therefore, the above three are on equal footing, and then, 
it may be interesting to investigate what results they give for 
each classical and quantal case.

\section{Classical treatment}

In \S 7 of (I), we sketched an idea how to treat the Hamiltonian obtained 
in classical and quantum framework. 
In this section, we apply this idea to the three classical cases 
discussed in \S 3. 
We treat the concrete case (\ref{2-8}) which is equivalent to the 
condition (\ref{2-7}) in the original fermion space. 
Further, in order to simplify various relations, we adopt $\hbar=1$ and 
$\epsilon=2$. 
Then, the functions $F(K)$ and $f(K)$ introduced in the form 
(I$\cdot$7$\cdot$1) is given in the following form: 
\beqn
& &H=F(K)-f(K)+2f(K)\left(\sin \psi/2\right)^2 \ . 
\qquad\qquad\qquad\qquad\qquad
\label{4-1}\\
& &F(K)=F_0+F_1(K) \ , 
\label{4-2}
\eeqn
\vspace{-0.5cm}
\bsub\label{4-3}
\beqn
& &F_0=-(2+G)\Omega \ , 
\label{4-3a}\\
& &F_1(K)=2(2-G\Omega)K+2GK^2 \ , 
\qquad\qquad\qquad\qquad\qquad
\qquad\ \ 
\label{4-3b}
\eeqn
\esub
\vspace{-0.5cm}
\begin{equation}\label{4-4}
f(K)=2G\sqrt{K(K+\delta)(\Omega-K)(\Omega+\Delta-K)}\ 
(=f(K;\delta,\Delta)) \ . 
\end{equation}
Here, $(\delta, \Delta)$ denote parameters for discriminating the three cases: 
\begin{eqnarray}\label{4-5}
& &(\ \delta=1 \ , \ \Delta=0\ ) \qquad {\rm for}\quad H(c_-) \ , \nonumber\\
& &(\ \delta=1 \ , \ \Delta=1\ ) \qquad {\rm for}\quad H(c^0) \ , \nonumber\\
& &(\ \delta=0 \ , \ \Delta=0\ ) \qquad {\rm for}\quad H(c) \ . 
\end{eqnarray}

Our picture is based on small amplitude oscillation around the energy 
minimum point. 
If the energy minimum point can be found in the region 
$0 < K < \Omega$ for $G\neq 0$, the form (\ref{4-1}) gives 
\begin{equation}\label{4-6}
\psi =0 \ .
\end{equation}
Then, our problem is reduced to finding a solution of the relation 
\begin{equation}\label{4-7}
F'(K)-f'(K)=0 \ , \qquad {\rm i.e.,}\qquad 
F_1'(K)-f'(K)=0 \ .
\end{equation}
Here, $F_1'(K)$ and $f'(K)$ are given as 
\beqn\label{4-8}
& &F_1'(K)=2(2-G\Omega)+4GK \ , \nonumber\\
& &f'(K)=G\biggl[\frac{2K+\delta}{\sqrt{K(K+\delta)}}
\sqrt{(\Omega-K)(\Omega+\Delta-K)} \nonumber\\
& &\qquad\qquad\qquad
-\sqrt{K(K+\delta)}\frac{2\Omega+\Delta-2K}{\sqrt{(\Omega-K)(\Omega+\Delta-K)}}
\biggl] \ .
\eeqn
%
\begin{figure}[t]
\begin{center}
\includegraphics[height=3.8cm]{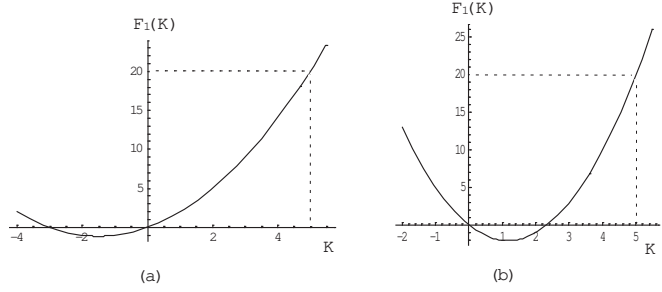}
\caption{The behavior of $F_1(K)$ is depicted in the case (a) $G\Omega<2$ 
($\Omega=5$ and $G=0.25$) 
and (b) $G\Omega >2$ ($\Omega=5$ and $G=0.75$). Here, $F_1'(0)=2(2-G\Omega)$. 
}
\label{fig:1}
\end{center}
\end{figure}
%
%
\begin{figure}[t]
\begin{center}
\includegraphics[height=3.8cm]{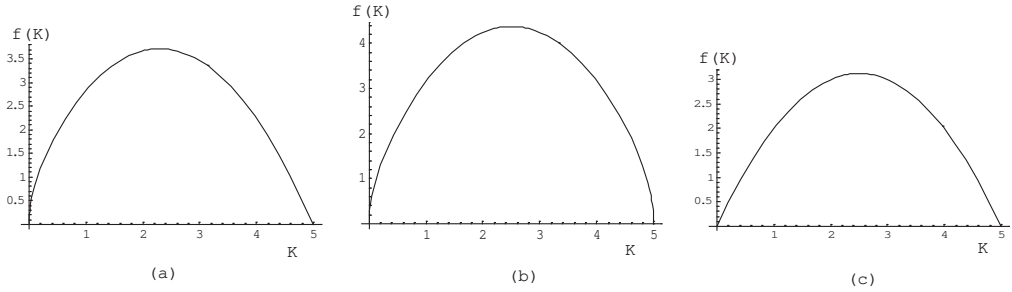}
\caption{The behavior of $f(K)$ is depicted in the cases of (a) $\delta=1,\ 
\Delta=0$ (case (i)), (b) $\delta=1,\ \Delta=1$ (case (ii)) and 
(a) $\delta=0,\ \Delta=0$ (case (iii)). In (a) and (b), $f'(0)$ goes to 
infinity, $f'(0)\rightarrow \infty$. On the other hand, in (c), 
$f'(0)=2G\Omega$. 
}
\label{fig:2}
\end{center}
\end{figure}
%
Figs.1 and 2 show that $F_1'(0)$ is finite and $f'(0)\rightarrow +\infty$ 
in the case $\delta=1$, and $(F_1(K)-f(K))$ is decreasing near 
$K=0$ and then, afterward, changes to increasing function until $K=\Omega$. 
Therefore, we can find the solution of the relation (\ref{4-7}) in the region 
$0 < K < \Omega$. 
On the other hand, in the case $(\delta=0,\ \Delta=0)$, if 
$F_1'(0)-f'(0) >0$, i.e., $G\Omega <1$, 
$(F_1(K)-f(K))$ is an increasing function in the region 
$0 < K < \Omega$ and we cannot find the solution of the relation 
(\ref{4-6}). 
In this case, $K=0$ gives the energy minimum point and the condition 
(\ref{4-6}) is not necessary. 
If $F_1'(0)-f'(0) <0$, i.e., $G\Omega >1$, $(F_1(K)-f(K))$ is 
decreasing near $K=0$ and then, afterward, changes to increasing function. 
Therefore, we can find the solution of the relation (\ref{4-6}). 
The above consideration gives the energy of the energy minimum point 
in the form 
\begin{equation}\label{4-9}
E_c(K_0;\delta,\Delta)=F_0+F_1(K_0)-f(K_0;\delta,\Delta) \ . 
\end{equation}
Here, for the cases $\delta=1$ and $\delta=0$ ($G\Omega >1$), $K_0$ 
is given as the solution of the relation (\ref{4-7}):
\begin{equation}\label{4-10}
F_1'(K_0)-f'(K_0;\delta,\Delta)=0 \ .
\end{equation}
For the case $\delta=0$ ($G\Omega <1$), $K_0$ is given as 
\begin{equation}\label{4-11}
K_0=0 \ .
\end{equation}

The frequency $\omega$ which characterizes the small amplitude oscillation 
around the energy minimum point is given in the form for the cases 
$\delta=1$ and $\delta=0$ ($G\Omega >1$) 
\begin{equation}\label{4-12}
\omega=\left[\left(F''(K_0)-f''(K_0)\right)f(K_0)\right]^{1/2} \ .
\end{equation}
This form can be found in the relation (I$\cdot$7$\cdot$5). 
Of course, the frequency (\ref{4-12}) depends on $K_0$, $\delta$ and 
$\Delta$. 
Then, denoting as $\omega(K_0;\delta,\Delta)$ and using the relation 
(\ref{4-10}), we have 
\beqn\label{4-13}
\omega(K_0;\delta,\Delta)&=&
\biggl[F_1'(K_0)^2+F_1''(K_0)f(K_0;\delta,\Delta) 
-(1/2)\left(f^2\right)''(K_0;\delta,\Delta)\biggl]^{1/2} \ . \qquad
\eeqn
Here, $(f^2)''(K;\delta,\Delta)$ denotes second derivative of 
$f(K;\delta,\Delta)^2$ for $K$. 
The frequency in the case $\delta=0$ ($G\Omega <1$) 
is also obtained. 
The method is discussed in the Appendix B. 
Thus, the Hamiltonian $H$ is expressed as 
\begin{equation}\label{4-14}
H=E_c(K_0;\delta,\Delta)+\omega(K_0;\delta,\Delta)d^*d \ .
\end{equation}
Here, $(d, d^*)$ denotes boson-type canonical variable.

The above general argument presents the following concrete expressions: \\
(i) For $\kket{c_-}$ ($\delta=1,\ \Delta=0$):
\beqn\label{4-15}
& &G^{-1}=(\Omega-2K_0)\cdot \frac{1}{2}\left(1+\sqrt{\frac{K_0}{K_0+1}}\right)
+\frac{\Omega-3K_0}{4\sqrt{K_0(K_0+1)}} \ , \nonumber\\
& &E_c(c_-)=-2\Omega+4K_0-G
\left[\Omega+2\sqrt{K_0}\left(\sqrt{K_0}+\sqrt{K_0+1}\right)(\Omega-K_0)
\right] \ , \nonumber\\
& &\omega(c_-)=2\biggl[(2-G\Omega)^2+4G(2-G\Omega)K_0 \nonumber\\
& &\qquad\qquad
-G^2\left(2K_0^2-3(2\Omega-1)K_0+\Omega(\Omega-2) 
-2\sqrt{K_0(K_0+1)}(\Omega-K_0)\right)\biggl]^{1/2} \ , \nonumber\\
& &
\eeqn
(ii) for $\kket{c^0}$ ($\delta=1,\ \Delta=1$):
\beqn\label{4-16}
& &G^{-1}=(\Omega-2K_0)\cdot \frac{1}{2}
\left(1+\sqrt{\frac{K_0(\Omega-K_0)}{(K_0+1)(\Omega-K_0+1)}}\right) \nonumber\\
& &\qquad\qquad\ 
+\frac{(\Omega-2K_0)(\Omega+1)}{4\sqrt{K_0(K_0+1)(\Omega-K_0)(\Omega-K_0+1)}} 
\ , \nonumber\\
& &E_c(c^0)=-2\Omega+4K_0\nonumber\\
& &\qquad\qquad \ 
-G\left[\Omega+2\sqrt{K_0(\Omega-K_0)}
\left(\sqrt{K_0(\Omega-K_0)}+\sqrt{(K_0+1)(\Omega-K_0+1)}\right)
\right] \ , \nonumber\\
& &\omega(c^0)=2\biggl[(2-G\Omega)^2+4G(2-G\Omega)K_0 \nonumber\\
& &\qquad\qquad
-G^2(2K_0^2-6\Omega K_0+(\Omega^2-\Omega-1) \nonumber\\
& &\qquad\qquad\qquad\quad
-2\sqrt{K_0(K_0+1)(\Omega-K_0)(\Omega+1-K_0)})\biggl]^{1/2} \ , 
\eeqn
(iii) for $\kket{c}$ ($\delta=0,\ \Delta=0,\ G\Omega <1$):
\bsub\label{4-17}
\beqn\label{4-17a}
& &K_0=0 \ , \nonumber\\
& &E_c(c)=-2\Omega-G\Omega\ , \nonumber\\
& &\omega(c)=4\sqrt{1-G\Omega}\ , 
\eeqn
(iii)' for $\kket{c}$ ($\delta=0,\ \Delta=0,\ G\Omega >1$):
\beqn\label{4-17b}
& &G^{-1}=\Omega-2K_0 \ , \nonumber\\
& &E_c(c)=G^{-1}-G\Omega(\Omega+1)\ , \nonumber\\
& &\omega(c)=2\sqrt{(G\Omega)^2-1}\ . 
\eeqn
\esub
%
\begin{figure}[t]
\begin{center}
\includegraphics[height=5.8cm]{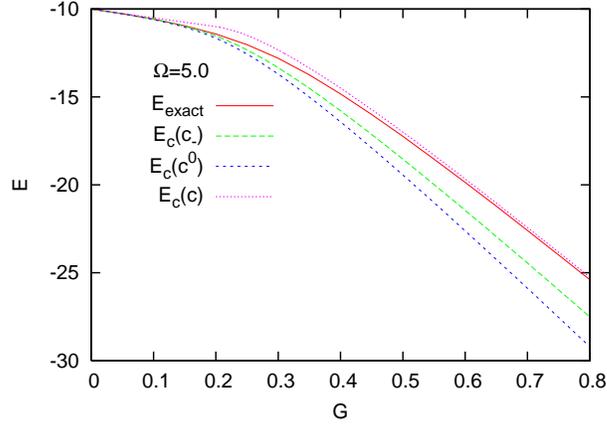}
\caption{The classical energies are depicted as a function of 
$G$ in the cases of 
(i) $\delta=1$ and $\Delta=0$ ($E_{c}(c_-)$), 
(ii) $\delta=1$ and $\Delta=1$ ($E_{c}(c^0)$) and 
(iii) $\delta=0$ and $\Delta=0$ ($E_{c}(c)$), where 
$\Omega=5$. 
Here, $E_{\rm exact}$ shows the exact eigenvalue of the Hamiltonian. 
}
\label{fig:3-0}
\end{center}
\end{figure}
%
%
Here, each $E_c$ is identical to the energy in the classical 
description of the 
two-level pairing model. In this paper, we have presented three different 
classical counterparts for the original system, which lead to the 
corresponding energies, $E_c(c_-)$, $E_c(c^0)$ and $E_c(c)$, respectively. 
Figure \ref{fig:3-0} shows them together 
with exact eigenvalue $E_{\rm exact}$. 
Then $E_c(c)$ reveals the energy obtained by using the 
$su(2)\otimes su(2)$-coherent state, that is, the Glauber 
coherent state, which corresponds to the energy obtained by 
the Hartree-Fock-Bogoliubov calculation in the original fermion system. 
It seems that the use of the $su(2)\otimes su(1,1)$-coherent 
state, $\kket{c_-}$, and the $su(1,1)\otimes su(1,1)$-coherent state, 
$\kket{c^0}$, comparatively give good results in comparison with 
the exact energy eigenvalue in the region of small force strength $G$. 
On the other hand, the $su(2)\otimes su(2)$-coherent state $\kket{c}$ 
gives rather good result in the region of large force strength $G$ 
in this stage. 
In the next section, we reinvestigate the ground state 
energy and the oscillation frequency 
around the static configuration in terms of the quantal description 
based on these three classical counterparts.

\section{Quantal treatment including quantum fluctuations}

In order to compare the energies derived in the above procedure 
with the exact ground-state energy in quantal description, 
we introduce the energy $E_q$ with quantum correction ${\cal E}$ as 
\begin{eqnarray}\label{4-18}
& &E_q=E_c+{\cal E} \ , \nonumber\\
& &{\cal E}=\frac{1}{2}\cdot
\left[f'(K_0)-\frac{f(K_0)}{K_0}-\frac{g'(K_0)}{g(K_0)}f(K_0)
+\omega\right] \ , 
\end{eqnarray}
where $g(K)\equiv g(K;\delta,\Delta)$ 
is defined through $f(K)$ in (\ref{4-4}) as 
\bsub\label{4-19}
\begin{eqnarray}
& &f(K;\delta,\Delta)=2G\sqrt{K(K+1)}(\Omega-K)
\sqrt{\frac{g(K;\delta,\Delta)}{g(K+1;\delta,\Delta)}}\ , 
\label{4-19a}\\
& &\ \ g(K;\delta=1,\Delta=0)=1\ , \nonumber\\
& &\ \ g(K;\delta=1,\Delta=1)=\Omega-K+1\ , \nonumber\\
& &\ \ g(K;\delta=0,\Delta=0)=K\ . 
\label{4-19b}
\eeqn
\esub
%
\begin{figure}[t]
\begin{center}
\includegraphics[height=5.8cm]{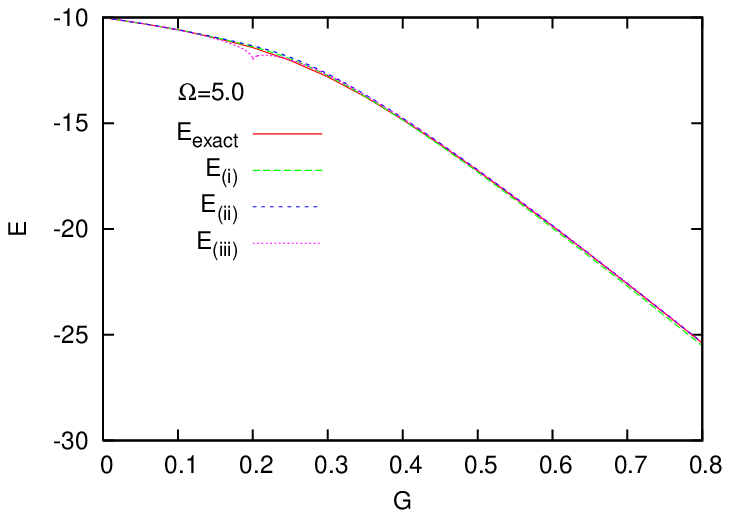}
\caption{The energies are depicted as a function of $G$ in the cases of 
(i) $\delta=1$ and $\Delta=0$ ($E_{\rm (i)}$), 
(ii) $\delta=1$ and $\Delta=1$ ($E_{\rm (ii)}$) and 
(iii) $\delta=0$ and $\Delta=0$ ($E_{\rm (iii)}$), where 
$\Omega=5$. 
Here, $E_{\rm exact}$ shows the exact eigenvalue of the Hamiltonian. 
}
\label{fig:3}
\end{center}
\end{figure}
%
%
\begin{figure}[t]
\begin{center}
\includegraphics[height=5.8cm]{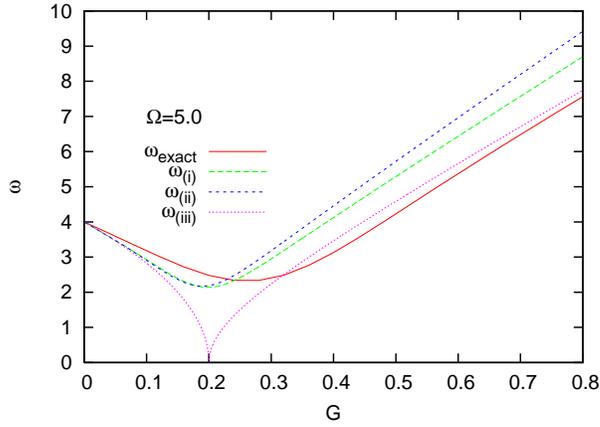}
\caption{The frequencies are depicted as a function of $G$ in the cases of 
(i) $\delta=1$ and $\Delta=0$ ($\omega_{\rm (i)}$), 
(ii) $\delta=1$ and $\Delta=1$ ($\omega_{\rm (ii)}$) and 
(iii) $\delta=0$ and $\Delta=0$ ($\omega_{\rm (iii)}$), where 
$\Omega=5$. 
Here, $\omega_{\rm exact}$ shows the exact result. 
}
\label{fig:4}
\end{center}
\end{figure}
%
In a form similar to the Hamiltonian (\ref{4-14}), we have the 
quantized Hamiltonian in the form 
\begin{equation}\label{5-2-add}
{\hat H}=E_q(K_0; \delta, \Delta)+\omega(K_0;\delta,\Delta){\hat d}^*{\hat d} 
\ .
\end{equation}
Here, it should be noted that the frequency $\omega$, which is given 
in the relation (\ref{4-13}), plays the role of the excitation energy in the 
present case. 
The derivation of (\ref{5-2-add}) with (\ref{4-18}) and  
the frequency $\omega$ 
is given in appendix B. 

Figure \ref{fig:3} shows the behavior of ${E}$ in the cases (i), 
(ii) and (iii), compared with the exact ground-state energy $E_{\rm exact}$, 
as a function of the strength of interaction $G$. The behavior is almost 
the same in each case except for case (iii) near $G=0.2$. 
In the case (iii), a dip appears at $G=0.2$, which corresponds to the 
phase transition point in the Hartree-Fock-Bogoliubov calculation, 
namely, $G\Omega=1$ with $\Omega=5$. 
The $su(2)\otimes su(2)$-coherent state $\kket{c}$ in (\ref{3-11}) 
in case (iii) is identical with 
the Glauber coherent state. This state corresponds to the Hartree-Fock 
and/or the Hartree-Fock-Bogoliubov state in the original fermion space. 
This behavior near $G=0.2$ in case (iii) is realized in the quasi-particle 
random phase approximation (QRPA)\cite{2}. Thus, it is seen that 
the energy correction ${\cal E}$ gives the ground state correlation 
in the random phase approximation. 
On the other hand, in the cases (i) and (ii), the dip structure 
does not appear.

Figure \ref{fig:4} shows the behavior 
of the frequency $\omega$ in the cases (i), 
(ii) and (iii), compared with the exact frequency $\omega_{\rm exact}$, 
which is exactly calculated for the excitation energy from the 
ground state. 
In the $su(2)\otimes su(2)$-coherent state $\kket{c}$, the dip 
structure appear at $G=0.2$, which is the same behavior as the 
QRPA calculation. However, by using the $su(2)\otimes su(1,1)$-coherent 
state $\kket{c_-}$ in case (i) and the $su(1,1)\otimes su(1,1)$-coherent 
state $\kket{c^0}$ in case (ii), this fictitious dip structure 
does not appear 
and the derived frequencies well reproduce the exact frequency. 
Especially, the result in the case (i) shows same behavior 
as the exact frequency in the region below the appearance of the dip in 
case (iii). 
However, the case (iii) based on the $su(2)\otimes su(2)$-coherent 
state approaches the exact frequency in the region 
of the large $G$.

\section{Concluding remarks}

We investigate the ground-state energy and the frequency of the small 
amplitude oscillation around the ground state in the two-level pairing model 
governed by the $su(2)\otimes su(2)$-algebra 
with the viewpoint of using the various states based on the $su(2)$- 
and the $su(1,1)$-algebras in the boson representations. 
Various classical descriptions are possible through the 
various states mentioned above, while the original fermionic quantum 
theory is unique. 

The $su(2)\otimes su(2)$-coherent state in the boson representation 
is identical with the Glauber coherent state, which corresponds to 
the BCS state (Hartree-Fock-Bogoliubov state) in the QRPA calculation. 
Then, the fictitious sharp phase transition point appears. 
However, taking into account the possibility of 
the various classical descriptions based on the various states, 
the fictitious sharp phase transition can be avoided. 
Thus, a reasonable approximation can be obtained by using other 
coherent states such as the $su(2)\otimes su(1,1)$- and $su(1,1)\otimes 
su(1,1)$-coherent states. 

Further numerical results will be reported in a subsequent paper, 
where $g(K)$ introduced in the relation (\ref{4-19a}) is adopted 
in a form different from those shown in the relation (\ref{4-19b}).

\section*{Acknowledgements} 
This work started when two of the authors (Y. T. and M. Y.) 
stayed at Coimbra in September 2005 and was completed when 
Y. T. stayed at Coimbra in September 2006. They would like to 
express their sincere thanks to Professor Jo\~ao da Provid\^encia, a 
co-author of this paper, for his invitation and warm hospitality. 
One of the authors (Y. T.) 
is partially supported by a Grant-in-Aid for Scientific Research 
(No.15740156 and No.18540278) 
from the Ministry of Education, Culture, Sports, Science and 
Technology of Japan. 


\appendix
\section{Proof of the result (\ref{4-17a})}

In this appendix, we show a proof of the result (\ref{4-17a}). 
For this aim, let us start with the classical Hamiltonian (\ref{3-12}), 
which was obtained under the Glauber coherent state, and 
treat the case in which the effect of $K$ is negligibly small 
compared with $\hbar \Omega$. 
Under this condition, by picking up linear terms for $K$, the Hamiltonian 
(\ref{3-12}) is approximated to 
\begin{equation}\label{b-1}
H=-\hbar(\epsilon +\hbar G\Omega)+2(\epsilon - \hbar G\Omega)K
-2\hbar G\Omega K \cos\psi\ .
\end{equation}
Of course, we consider the case $L=M=\hbar\Omega/2$ and $T=\hbar/2$. 
Since $(\psi, K)$ is a set of canonical variables, we have the following 
Hamilton equation of motion: 
\bsub\label{b-2}
\beqn
& &{\dot \psi}=\frac{\partial H}{\partial K}=2(\epsilon-\hbar G\Omega)
-2\hbar G\Omega \cos\psi \ , 
\label{b-2a}\\
& &{\dot K}=-\frac{\partial H}{\partial \psi}=
-2\hbar G\Omega K\sin\psi \ . 
\label{b-2b}
\eeqn
\esub
We can solve Eq.(\ref{b-2}) directly. 
But, it may be much more transparent to solve it by introducing the 
quantities 
\begin{equation}\label{b-3}
T_x=K\cos\psi \ , \qquad 
T_y=K\sin\psi \ , \qquad
T_z=K \ . 
\end{equation}
These are constrained by the condition 
\begin{equation}\label{b-4}
T_z^2-(T_x^2+T_y^2)=0 \ .
\end{equation}
We describe the present system in terms of $(T_x, T_y, T_z)$. 
With the use of the relation (\ref{b-3}), the time-derivative 
$({\dot T}_x,{\dot T}_y,{\dot T}_z)$ is given in the form 
\bsub\label{b-5}
\beqn
& &{\dot T}_x=-2(\epsilon-\hbar G\Omega)T_y \ , 
\label{b-5a}\\
& &{\dot T}_y=2(\epsilon-\hbar G\Omega)T_x-2\hbar G \Omega T_z \ , 
\label{b-5b}\\
& &{\dot T}_z=-2\hbar G\Omega T_y \ . 
\label{b-5c}
\eeqn
\esub
The relation (\ref{b-5}) gives us 
\begin{equation}\label{b-6}
{\ddot T}_y=-(2\epsilon)^2\left(1-\frac{2\hbar G\Omega}{\epsilon}\right)T_y \ .
\end{equation}
Since $T_y$ should be finite, the solution of Eq.(\ref{b-6}) is 
meaningful in the case 
\begin{equation}\label{b-7}
1-\frac{2\hbar G\Omega}{\epsilon} >0\ .
\end{equation}
Then, we define the frequency $\omega$ in the form 
\begin{equation}\label{b-8}
\omega=2\epsilon\sqrt{1-\frac{2\hbar G\Omega}{\epsilon}} \ .
\end{equation}
It may be clear that $T_y$ shows a harmonic oscillator type behavior and the 
relation (\ref{b-4}) and (\ref{b-5}) give 
\bsub\label{b-9}
\beqn
& &T_x=A\left[\frac{2(\epsilon-\hbar G\Omega)}{\omega}\cos(\omega t+\alpha)
+\frac{2\hbar G\Omega}{\omega}\right] \ , 
\label{b-9a}\\
& &T_y=A\sin(\omega t+\alpha) \ , 
\label{b-9b}\\
& &T_z=A\left[\frac{2\hbar G\Omega}{\omega}\cos(\omega t+\alpha)
+\frac{2(\epsilon-\hbar G\Omega)}{\omega}\right] \ . 
\label{b-9c}
\eeqn
\esub
Here, $A$ and $\alpha$ denote constants to be determined by the initial 
condition. 
The Hamiltonian (\ref{b-1}) reduces to 
\begin{equation}\label{b-10}
H=-\hbar(\epsilon+\hbar G)\Omega+\omega A \ .
\end{equation}
If we put $\hbar=1$ and $\epsilon=2$, the above result obtains 
the form (\ref{4-17a}). 
Later, the meaning of $\omega$ will be discussed. 

If we disregard the starting condition $K\ll \hbar \Omega$, the present 
result can be accepted in the region 
$2\hbar G\Omega/\epsilon <1$ and any value of $A$. 
If we keep in mind the starting condition, some discussions are necessary. 
Since $-1 \leq \cos\theta \leq 1$ and $T_z=K$, the form (\ref{b-9c}) 
gives us 
\begin{equation}\label{b-11}
A\left(\frac{\omega}{2\epsilon}\right) \leq K \leq 
A\left(\frac{2\epsilon}{\omega}\right) \ .
\end{equation}
Then, combining $K \ll \hbar\Omega$ with the relation (\ref{b-11}), 
we have $A(2\epsilon/\omega)\ll \hbar \Omega$, i.e., 
\begin{equation}\label{b-12}
A \ll \hbar\Omega\cdot\frac{\omega}{2\epsilon} \ . 
\end{equation}
If $\omega$ becomes smaller, the value of $A$ may also become 
smaller. 

It may be interesting to see that $(T_x,T_y,T_z)$ obeys the 
$su(1,1)$-algebra classically. 
This can be seen in the relation 
\begin{equation}\label{b-13}
[\ T_+\ , \ T_-\ ]_P=2iT_0 \ , \qquad
[\ T_0\ , \ T_\pm\ ]_P=\mp iT_\pm \ . 
\end{equation}
Here, $[\ , \ ]_P$ denotes the Poisson bracket and 
$(T_{\pm,0})$ is defined as 
\begin{equation}\label{b-14}
T_\pm=T_x \pm iT_y \ , \qquad T_0=T_z \ . 
\end{equation}
The relation (\ref{b-4}) is rewritten as 
\begin{equation}\label{b-15}
T_0^2-T_+T_-=0 \ .
\end{equation}

\section{The derivation of quantal Hamiltonian with the disguised form}

In this appendix, we give the quantal Hamiltonian describing 
the small oscillation around the energy minimum state in a slightly 
disguised form for the original Hamiltonian (\ref{3-1}). 

A part of the Hamiltonian, ${\hat H}_1$, is recast into 
\beqn\label{a-1}
{\hat H}_1&=&-G\left[\sqrt{\hbar}{\hat c}^*\cdot\sqrt{2T+{\hat K}}
\sqrt{2M-{\hat K}}\sqrt{2L-{\hat K}} + {\rm h.c.}\right] \nonumber\\
&=&-G\Biggl[\Biggl(
\sqrt{g({\hat K})}\left(\sqrt{\hbar}{\hat c}^*-\sqrt{{\hat K}}\right)
\sqrt{2T+{\hat K}}
\sqrt{2M-{\hat K}}\sqrt{2L-{\hat K}}\frac{1}{\sqrt{g({\hat K}+\hbar)}} 
\nonumber\\
& &\ \ \ \ \ \  + \sqrt{{\hat K}}\sqrt{2T+{\hat K}}
\sqrt{2M-{\hat K}}\sqrt{2L-{\hat K}}
\sqrt{\frac{g({\hat K})}{g({\hat K}+\hbar)}}\Biggl)
+ {\rm h.c.}\Biggl] \ , 
\eeqn
where ${\hat K}=\hbar{\hat c}^*{\hat c}$ and $g({\hat K})$ is an arbitrary 
function with respect to ${\hat K}$. Here, h.c. represents the Hermite 
conjugate. 
Thus, the Hamiltonian (\ref{3-1}) can be expressed as 
\beqn\label{a-2}
& &{\hat H}=F_0+F_1({\hat K})-f({\hat K}) \nonumber\\
& &\qquad
-\frac{1}{2}\biggl[2G\sqrt{g({\hat K})}(\sqrt{\hbar}{\hat c}^*-
\sqrt{{\hat K}})\sqrt{2T+{\hat K}}\sqrt{2M-{\hat K}}\sqrt{2L-{\hat K}}
\frac{1}{\sqrt{g({\hat K}+\hbar)}} + {\rm h.c.} \biggl] \ , 
\nonumber\\
& & F_0=-\left[\epsilon -G(L+M-(T-\hbar/2))+4GTM\right] \ , \nonumber\\
& &F_1({\hat K})=2\left[\epsilon-G(L+M-(T-\hbar/2))\right]{\hat K}
+2G{\hat K}^2 \ , \nonumber\\
& &f({\hat K})=2G\sqrt{{\hat K}}\sqrt{2T+{\hat K}}\sqrt{2M-{\hat K}}
\sqrt{2L-{\hat K}}
\sqrt{\frac{g({\hat K})}{g({\hat K}+\hbar)}}  \ .
\eeqn
Here, let us focus on the last bracket in the Hamiltonian ${\hat H}$. 
From (I$\cdot$7$\cdot$11), we know that 
$\sqrt{\hbar}{\hat c}^*-
\sqrt{{\hat K}}=i\sqrt{\hbar}({\hat \gamma}^*+{\hat \gamma})/2
-((\sqrt{\hbar}({\hat \gamma}^*+{\hat \gamma}))^2-2\hbar)/8\sqrt{K_0}$. 
Then, substituting the above relation into the last term and 
expanding the terms up to the second order of $({\hat \gamma}, 
{\hat \gamma}^*)$, we obtain 
\beqn\label{a-3}
{\hat H}&=&F_0+F_1(K_0)-f(K_0)+(F_1'(K_0)-f'(K_0))\delta {\hat K}
+\frac{1}{2}(F_1''(K_0)-f''(K_0))
(\delta {\hat K})^2 
\nonumber\\
& &
+\frac{1}{2}f(K_0)\cdot\left(
\frac{1}{2\sqrt{K_0}}\sqrt{\hbar}({\hat \gamma}^*
+{\hat \gamma})\right)^2 
+\frac{\hbar}{2}\left(f'(K_0)-\frac{f(K_0)}{K_0}
-\frac{g'(K_0)}{g(K_0)}f(K_0)\right) \ . \nonumber\\
& &
\eeqn
Here, we have introduced $\delta {\hat K}={\hat K}-K_0=
\sqrt{K_0}\cdot i\sqrt{\hbar}({\hat \gamma}^*-{\hat \gamma})
+\hbar{\hat \gamma}^*{\hat \gamma}$ in (I$\cdot$7$\cdot$8) and 
used the commutation relation 
$[\ \delta{\hat K}\ , \ i\sqrt{\hbar}({\hat \gamma}^*+{\hat \gamma})/2\ ]
=\hbar\sqrt{K_0}$. 
Then, $K_0$ is determined by eliminating the linear term of $\delta{\hat K}$ 
as $F_1'(K_0)-f'(K_0)=0$ in Eq. (\ref{4-7}). Further, the above mentioned 
commutation relation can be expressed as 
$[\ \delta{\hat K}\ , \ \sqrt{\hbar}({\hat \gamma}^*+{\hat \gamma})/2\sqrt{K_0}
\ ]=-i\hbar$. 
Thus, the above Hamiltonian (\ref{a-3}) is easily diagonalized 
by means of newly introduced operator $({\hat d}, {\hat d}^*)$, which 
is written as 
${\hat d}=u \sqrt{\hbar}({\hat \gamma}^*+{\hat \gamma})/2\sqrt{K_0}+
v\delta{\hat K}$ with $u$-$v$ factor satisfying $u^2-|v|^2=1$. 
We finally obtain the Hamiltonian as 
\beqn
{\hat H}&=&F_0+F_1(K_0)-f(K_0)+\hbar\omega{\hat d}^*{\hat d} \nonumber\\
& &+\frac{\hbar}{2}\left(f'(K_0)-\frac{f(K_0)}{K_0}
-\frac{g'(K_0)}{g(K_0)}f(K_0)+\omega\right) \ , 
\label{a-4}\\
\omega&=&\sqrt{(F_1''(K_0)-f''(K_0))f(K_0)} \ . 
\label{a-5}
\eeqn
In this paper, since we discuss the case $L=M=(\hbar/2)\Omega$ and 
$T=\hbar/2$ in Eq.(\ref{2-8}) and the parameter $\epsilon=2$ in 
the natural unit $\hbar=1$, 
we obtain (\ref{4-3}) and (\ref{4-19}).

\end{document}